# Unexpected catalytic activity of nanorippled graphene

P. Z. Sun[1,2,3], W. Q. Xiong[4], A. Bera[1], I. Timokhin[1], Z. F. Wu[1,2], A. Mishchenko[1], M. C. Sellers[1], B. L. Liu[5], H. M. Cheng[5], E. Janzen[6], J. H. Edgar[6], I. V. Grigorieva[1,2], S. J. Yuan[4], A. K. Geim[1,2]

[1]Department of Physics and Astronomy, University of Manchester, Manchester, M13 9PL, UK
[2]National Graphene Institute, University of Manchester, Manchester, M13 9PL, UK
[3]Institute of Applied Physics & Materials Engineering, University of Macau, Taipa 999078, China
[4]School of Physics & Technology, Wuhan University, Wuhan, China
[5]Graphene Center, Institute of Materials Research, Tsinghua University, Shenzhen, 518055, China
[6]Tim Taylor Department of Chemical Engineering, Kansas State University, Manhattan, KS 66506, USA

*Graphite is one of the most chemically inert materials. Its elementary constituent, monolayer graphene, is generally expected to inherit most of the parent material's properties including chemical inertness. Here we show that, unlike graphite, defect-free monolayer graphene exhibits a strong activity with respect to splitting molecular hydrogen, which is comparable to that of metallic and other known catalysts for this reaction. We attribute the unexpected catalytic activity to surface corrugations (nanoscale ripples), a conclusion supported by theory. Nanoripples are likely to play a role in other chemical reactions involving graphene and, because nanorippling is inherent to atomically thin crystals, can be important for two dimensional materials in general.*

Despite inheriting chemical inertness from graphite and being highly stable in air, graphene has widely (and perhaps somewhat surprisingly) been considered for the use in catalysis (1-7). The primary reason is the huge surface area of graphene such that it can act as an efficient support and delivery platform for dispersed catalytically-active nanoparticles and functional groups (1-5). Another reason is that graphene itself can be modified relatively easily to provide a high density of vacancies, substitutional dopants, edges and other atomic-scale defects, which can potentially provide catalytically active sites (1-7). However, it was also noticed occasionally that even defect-free graphene monocrystals (obtained by mechanical exfoliations) could be more chemically reactive than bulk graphite. For example, monolayer graphene reacts with oxygen more easily than few-layer graphene and graphite (8). Also, wrinkles and strained regions of defect-free graphene were reported to accelerate its functionalization with certain chemicals (9-13). Despite those few observations, it remains unknown if pristine graphene (defect-free monolayers without wrinkles and macroscopic strain) is chemically inert, similar to graphite. One recent finding indicates that this might not be the case (14). Indeed, defect-free graphene membranes allow discernable permeation of molecular hydrogen, despite being completely impermeable to smaller and generally more permeating helium atoms (14). To explain this conundrum, it was speculated that, unlike graphite, graphene could split molecular hydrogen into two protons (hydrogen atoms adsorbed on graphene) and then those subatomic particles permeated through the graphene lattice, a feat essentially impossible for any gas under ambient conditions (15, 16). Although rather convoluted, this explanation was supported by theoretical predictions that strongly curved regions of graphene could dissociate molecular hydrogen (17-19) and the experimental fact that nominally flat 2D membranes exhibit ubiquitous nanoscale ripples with substantial strain and curvature (20-23).

In this report, we provide clear evidence for hydrogen dissociation on pristine graphene using three complementary sets of experiments. First, we have compared hydrogen permeation through graphene with that through so-called 'white graphene', monolayers of hexagonal boron nitride (hBN). This comparison is critical because the two 2D materials are structurally similar but have different electronic spectra which makes hydrogen dissociation possible only on graphene's nanoripples, according to theory (18, 19). No sign of hydrogen permeation through hBN monolayers could be found experimentally, despite the latter being hundred times more proton-conductive than graphene (15, 16). The provided side-by-side comparison between these two rather similar monolayer materials also rules out any possible artefacts (e.g., influence of accidental defects). Second, we have used Raman spectroscopy to monitor the development of the D peak that appears if molecular hydrogen is split into atoms and those bind to the carbon lattice (24, 25). We find that the D peak appears for graphene placed onto a silicon-oxide substrate that is relatively rough and allows nanoripples, whereas no sign of hydrogen adsorption could be detected for either ripple-free (atomically flat) graphene or bulk graphite's surface. These observations corroborate that roughness plays a critical role in hydrogen dissociation and emphasize further the difference between reactivities of defect-free graphene and graphite surfaces. Third, using a mixture of hydrogen and deuterium gases, we show that graphene acts as a powerful catalyst converting $H_2$ and $D_2$ into HD, in contrast to graphite and other carbon-based materials under the same conditions. Graphene's catalytic efficiency per gram exceeds by far that of traditional catalysts for the reaction.

**Hydrogen permeation through graphene and hBN monolayers**

We used the same setup and procedures as described in detail previously (14). Briefly, micrometer-size containers were fabricated from monocrystals of graphite and sealed with monolayers of either graphene or hBN (Figs. 1*A*, *B*). The microcontainers were placed in a gas chamber filled with, e.g., helium. If defects were present or the sealing was imperfect, the membranes started bulging because of a gradually increasing pressure inside (*SI Appendix*, Fig. S1). The bulging could be monitored by atomic force microscopy (AFM), and changes in the membrane position $\delta$ with time (Figs. 1*B*, *C*) could be translated directly into permeation rates (14, 26). This technique is remarkably sensitive, allowing detection of even a single vacancy if present in graphene membranes (26). In the absence of defects, our microcontainers were completely impermeable to any inert gas, within an experimental accuracy of a few atoms per hour (leak rate < $10^9$ $s^{-1}$ $m^{-2}$) (14).

We placed such helium-proof microcontainers inside a chamber filled with hydrogen under ambient conditions and found that graphene membranes slowly but steadily bulged out over days (Figs. 1*C*, *D*). The measured leak rate of ~$2\times10^{10}$ $s^{-1}$ $m^{-2}$ (Figs. 1*D*, *E*) agreed with our earlier report (14). As the energy barrier for helium permeation through graphene is predicted to be several eV and even higher for molecular hydrogen, the only possibility for hydrogen to pass through defect-free graphene is as protons because the latter face a relatively low barrier of about 1 eV (for the case of graphene monolayers) (16). Accordingly, the hydrogen-gas permeation has been explained by a two-stage mechanism: first, molecular hydrogen dissociates on graphene (on top of nanoripples as suggested by theory) (14, 18, 19) and then the resulting hydrogen adatoms, indistinguishable from adsorbed protons, flip to the other side of graphene by overcoming the same barrier as measured for proton transport (15, 16). This part of our report repeats the earlier experiments and is provided here only to make a clear comparison between hBN and graphene monolayers (see below).

In stark contrast to graphene, similar containers but sealed with hBN monolayers did not bulge in either helium or hydrogen atmosphere (that is, exhibited no sign of hydrogen permeation; Figs. 1*C*, *D*). If hBN were to dissociate hydrogen (the first stage of the above mechanism), then hydrogen would permeate much more quickly through hBN than graphene because the former is two orders of magnitude more proton-transparent (15). The absence of any discernable hydrogen permeation proves that, unlike graphene, hBN is not active for splitting $H_2$, in agreement with our density functional theory calculations (*SI Appendix*, Figs. S2-5).

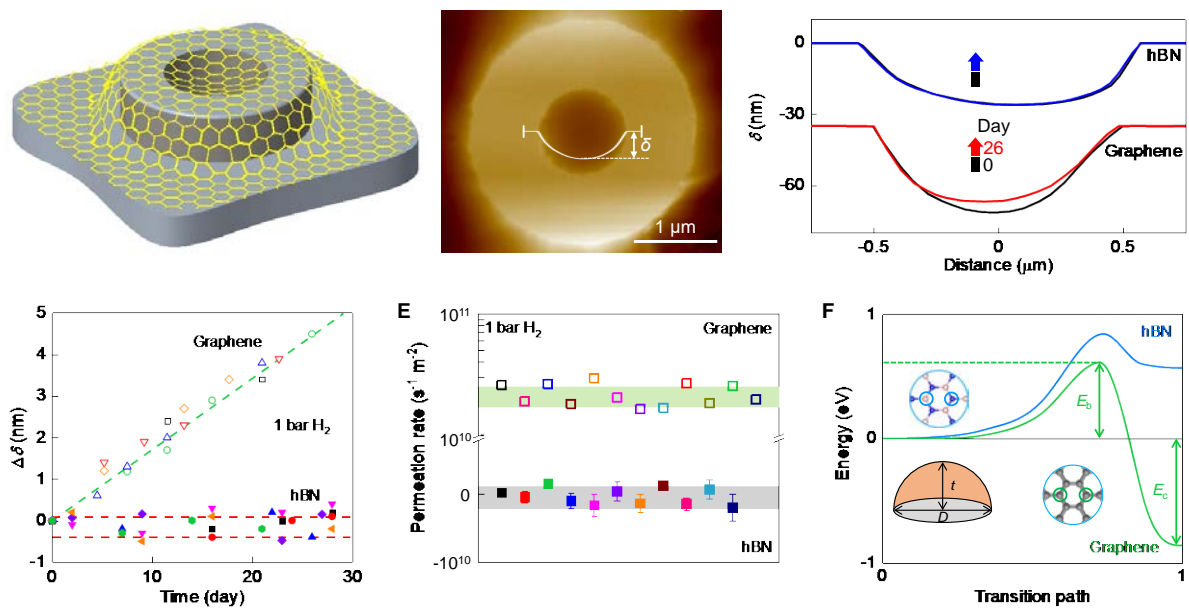

**Fig. 1.** Hydrogen transport through graphene and hBN membranes. (*A*) Schematic of the microcontainers. (*B*) AFM image of one of our microcontainers sealed with an hBN monolayer. The white curve shows the height profile along the membrane center. (*C*) Height profiles for two containers sealed with graphene and hBN before and after their storage in hydrogen for a few weeks (color-coded curves). $T$ = 295 ± 3 K; $P$ = 1 bar. Note the different scales for the x- and y- axes, which serve to show changes in $\delta$ and exaggerate the inward curving of 2D membranes. This sagging is caused by graphene's adhesion to inner walls of the microcontainers. The macroscopic curvature due to sagging was less than 4% and insufficient to cause hydrogen dissociation, according to theory (18, 19). Nanoscale rippling provides considerably higher curvature and strain (20, 22, 23). (*D*) Changes in the height $\Delta\delta$ for hBN (solid symbols) and graphene (open) membranes measured over one month. Each symbol in (*D*, *E*) denotes a different microcontainer with more than 10 tested for each material. The red dashed lines outline the full range of $\Delta\delta$ for the hBN membrane. Green line: best linear fit for the graphene data shown by the green symbols. (*E*) Permeation rates evaluated from the $\Delta\delta$ measurements. Note a break in the y-axis in (*E*): the scale is linear below $10^{10}$ and logarithmic above. Error bars: SD for the linear fits of $\Delta\delta$ as a function of time in (*D*); shown only if larger than the symbols. Shaded areas: overall SD for the graphene and hBN measurements. (*F*) Energy profiles for dissociation of $H_2$ on graphene and hBN ripples for the case of two hydrogen atoms being adsorbed at the central positions as shown schematically in the insets. Other adsorption positions are less favorable for hydrogen splitting as discussed in *SI Appendix*. The curvature $t/D$ (low-left inset) is 12% for both curves in (*F*).

These calculations show a clear difference between graphene and hBN. If their monolayers are flat, the dissociation barrier $E_b$ (calculated as the maximum energy at the transition state, as illustrated in Fig. 1*F*) is very high (> 3 eV) for both cases. Ripples with considerable curvature $t/D$ can reduce the barrier to less than 1 eV (where $t$ is the height of a ripple and $D$ its lateral dimension; inset of Fig. 1*F*). However, hydrogen dissociation can be energetically favorable only for graphene (that is, it becomes an exothermic reaction characterized by the chemisorption energy $E_c$ <0, as shown in Fig. 1*F*). This requires curvatures above ~10% (Fig. 1*F*; Fig. S2). Such strongly-curved nanoripples are routine for graphene membranes. Their rippling was extensively studied by transmission electron microscopy (20) and scanning tunneling microscopy (22, 23) and attributed to both local strain and thermal fluctuations (20, 21, 23). In the case of hBN monolayers, hydrogen dissociation is calculated to remain energetically unfavorable ($E_c$ > 0) for all realistic curvatures and strains (<15%) that can be supported by 2D crystals without breaking (Fig. 1*F*, Figs. S2-4).

The above results suggest a dramatic difference in chemical and catalytic activity of graphene and hBN monolayers, at least with respect to hydrogen. However, with only hundred molecules per hour permeating inside graphene microcontainers (14), neither experiment nor theory provides any indication whether graphene is a good or bad catalyst for hydrogen dissociation.

**Role of graphene's non-flatness**

The described mechanism of hydrogen dissociation by graphene nanoripples implies that, if placed in a hydrogen atmosphere, graphene's surface should contain a certain amount of adsorbed hydrogen atoms. Their binding to carbon atoms of the crystal lattice (C-H bonds) is strong ($sp^3$-type) (18, 19), especially if hydrogen atoms occupy neighboring carbon sites, like in the case of a fully-hydrogenated graphene derivative, graphane (24, 25). Such $sp^3$ adatoms should give rise to the D peak in graphene's Raman spectra. Accordingly, this peak can be used as a measure of graphene's reactivity with respect to hydrogen dissociation.

At room temperature (*T*), the coverage of graphene with hydrogen adatoms is expected to be extremely sparse (~100 hydrogen molecules permeate through micrometer-scale membranes during an entire hour) (14) and, indeed, we did not observe the appearance of any discernable D peak for monolayer graphene placed in a hydrogen atmosphere for many days. To increase the coverage, graphene can be heated to help overcome a finite energy barrier required for hydrogen dissociation (Fig. 1*F*). Moreover, thermal fluctuations at higher *T* can also generate dynamic nanoripples with high curvature (21, 23). With these considerations in mind, we heated graphene monolayers to a certain *T* in pure $H_2$ and compared their Raman spectra with similarly treated graphene but in He. For these measurements, graphene was exfoliated directly onto a silicon-oxide wafer because suspended graphene cannot withstand high *T* and breaks down, presumably due to induced mechanical strain (14). Note that the silicon-oxide surface supporting graphene was relatively rough with a root-mean-square roughness of ~0.5 nm (Fig. S6). As shown in Fig. 2*A*, graphene crystals heat-treated in hydrogen and then cooled down to room *T* exhibited a pronounced D peak, indicating the appearance of numerous $sp^3$ defects. In contrast, similar graphene samples but heated in a helium atmosphere showed no sign of the D peak, proving qualitatively that graphene could react only with hydrogen.

Next, we used the D peak to assess a role of graphene's surface roughness in the hydrogenation process. To this end, the above Raman results were compared with those obtained for similarly exfoliated graphene monolayers but placed onto an atomically flat surface of graphite (*SI Appendix*, Fig. S6). After exposing the flat graphene to 600 °C in pure $H_2$ for several hours, no sign of the D peak could be discerned in the Raman spectra, in clear contrast to the case of non-flat graphene (Fig. 2*A*). Neighboring graphite surfaces also showed no D peak. Note that, despite different thicknesses, graphene monolayers and bulk graphite exhibit similar intensities of the G peak (Fig. 2*A*), in agreement with the previous studies (27). Accordingly, if the D peak even with a 100 times smaller intensity were present for graphene placed on graphite, our sensitivity would certainly allow its detection (inset of Fig. 2*A*). In another control experiment, we prepared graphene with intentionally introduced $sp^3$ defects (e.g., tears) and again placed the samples on atomically flat graphite. In this case, the D peak was clearly seen, proving that we would easily see hydrogen adatoms if they were also adsorbed on flat graphene.

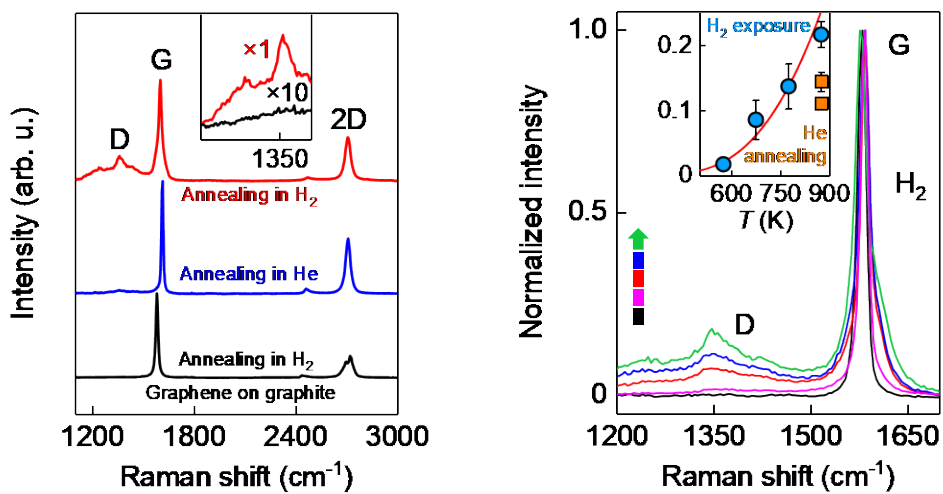

**Fig. 2.** Raman spectroscopy of graphene exposed to hydrogen. (*A*) Raman spectra of monolayer graphene after its exposure to 1-bar hydrogen at 600 °C for 2 hours (red curve). Blue curve: same but for helium. Black curve: same measurements but for atomically-flat graphene placed on graphite without alignment and exposed to hydrogen at 600 °C for 4 h. The spectra were taken at room *T* and, for clarity, are shifted vertically. All spectroscopy parameters were exactly the same: wavelength, 514.5 nm; power, ∼1.7 mW; spot size, ∼2 μm; acquisition time for each curve, 1 h. The inset magnifies the D peak region on the red and black curves with the intensity for the black curve being amplified 10 times. (*B*) Raman spectra of graphene in hydrogen taken in situ with increasing *T* (color coded). Inset: *T* dependence of the D peak intensity relative to that of the G peak (circles). Squares: hydrogen-induced D peak decreases after 4 and 8 hours of annealing at 600 °C in a helium atmosphere (top and bottom squares, respectively). The spectra were taken from the same spots away from graphene edges. Error bars: SD for at least 3 different spots used in the Raman measurements. Solid curve: guide to the eye showing the Arrhenius dependence with an activation energy of 0.4 eV.

Consecutive annealing of hydrogenated (non-flat) graphene in He or vacuum reduced the D peak, although a notable hump remained in the relevant spectral region even after several days at 600 °C (inset of Fig. 2*B*, Fig. S7). This hump is likely to be due to small regions of thermally stable graphane, which is formed if graphene is covered with atomic hydrogen on both sides (24). To quantify the observed hydrogenation process, Fig. 2*B* shows consecutive Raman spectra taken inside a 1-bar hydrogen atmosphere with increasing *T*. The D peak gradually grows with increasing *T*, and this dependence allows an estimate for the hydrogen-reaction activation energy as ~0.4 eV.

The results reported in this section suggest that non-flatness is essential for graphene's reactivity and that the activation energy for hydrogen dissociation is relatively small. However, they still provide no indication about graphene's efficiency as a catalyst. The latter is assessed in the next section.

**Enhanced hydrogen isotope exchange**

Annealing hydrogenated graphene in helium or vacuum shows that hydrogen adatoms desorb from graphene's surface (inset of Fig. 2*B*) and then presumably recombine into molecular hydrogen. This agrees with the proposed mechanism of hydrogen-gas permeation through graphene membranes, which involves recombination of permeating protons (atomic hydrogen) inside microcontainers (14-16). If instead of pure $H_2$, graphene is exposed to a mixture of $H_2$ and $D_2$, one should expect hydrogen and deuterium adatoms to recombine at random, forming an HD gas (in addition to $H_2$ and $D_2$). Therefore, detection of HD would provide an unequivocal proof for hydrogen dissociation on graphene and, using this reaction, graphene could also be benchmarked against known catalysts.

Because individual graphene crystals generate such a minute amount of HD, it is impossible to detect it by mass spectrometry. Accordingly, we chose to use a graphene powder for these experiments. It was obtained through careful reduction of graphene oxide, which induced little defects in the basal plane (*ACS Material*; Fig. S8). The powder was well-characterized and contained highly corrugated and mostly isolated (non-restacked) monolayers with the measured specific surface area of ~1,000 $m^2\ g^{-1}$ (28). A quartz tube was fully filled with this powder, and a mixture of hydrogen and deuterium gases ($P_{H2} = P_{D2} \approx 0.5$ bar) was then added (*SI Appendix*). After a certain exposure time, the mixed gas was analyzed by mass spectroscopy to determine the resulting HD concentration, $\rho_{HD}$. At room *T*, we could not detect any HD either in the presence of graphene or without it (within our experimental accuracy, better than 0.2%, being limited by the background HD present inside the commercially supplied $D_2$). To accelerate the reaction of $H_2$ with $D_2$, the gas mixture was heated up. At 600 °C, the monolayer graphene powder led to formation of HD, and its concentration increased as a function of time saturating at ~15% after approximately 5 hours (inset of Fig. 3*A*). No HD could be detected in the absence of graphene. The dissociation-recombination reaction $H_2$ + $D_2 \leftrightarrow$ 2HD can result in $\rho_{HD}$ up to 50%, if $H_2$ and $D_2$ molecules are fully split and then recombined stochastically into HD, $H_2$ and $D_2$ products. The observed saturation below 50% can be attributed to 'poisoning', a standard occurrence for catalysts (Fig. 3*B* and Fig. S9). For graphene, the poisoning is probably caused by formation of chemically inert regions of graphane (24, 25), a conclusion also supported by the Raman behavior discussed in the previous section. Furthermore, the inset of Fig. 3*A* shows that during initial stages the HD reaction proceeded at constant production rates, $d\rho_{HD}/dt$. By plotting these rates as a function of *T*, we have found that they follow the Arrhenius behavior (Fig. 3*A*). The fit yields the activation energy of ~0.4 eV, which agrees well with our estimate obtained from the Raman spectroscopy (Fig. 2*B*).

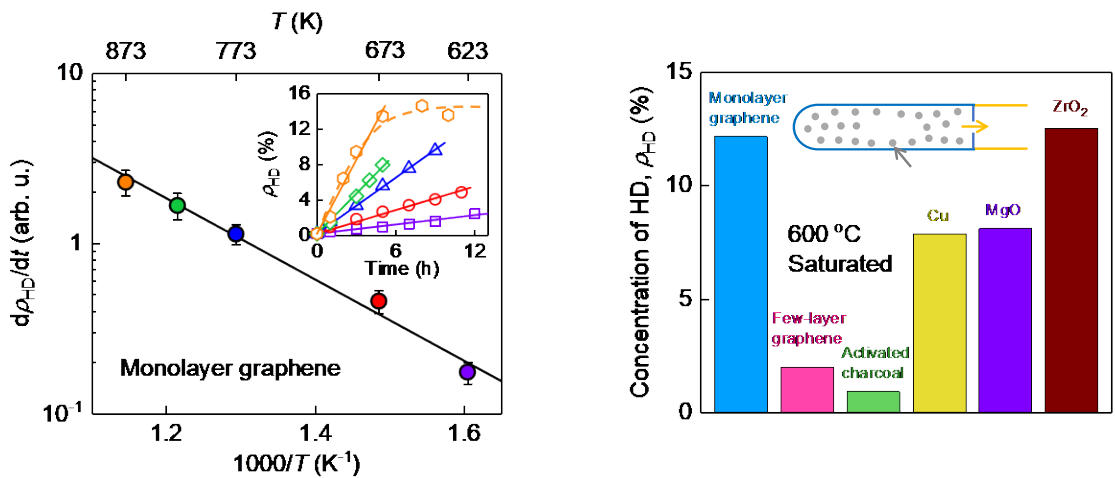

**Fig. 3.** Hydrogen-deuterium exchange catalyzed by graphene. (*A*) Temperature dependence of HD production rates for a monolayer graphene powder. Symbols: experimental data at different *T* with error bars indicating SD. Solid curve: best exponential fit, yielding the activation energy of ~0.4 eV. Inset: $\rho_{HD}$ as a function of time *t* at different *T* (color coding for *T* as in the main panel). Solid lines: linear fits for the initial stage of HD production. Dashed curve: guide to the eye. Our accuracy of determining the HD concentrations was ±2 % as found in repeated measurements under the same conditions. (*B*) Final concentration of HD after annealing the $H_2/D_2$ mixtures using various catalysts and reaching saturation in $\rho_{HD}$ as a function of time (5 hours at 600 °C in all cases). Inset: schematic of the experimental setup where MS refers to 'mass spectrometer'.

As a reference, we carried out the same measurements using known catalysts for the HD reaction, namely, zirconia, MgO and Cu (29, 30). They were also in the powder form as detailed in *SI Appendix*. The HD production for these catalysts exhibited time dependences qualitatively similar to that of the monolayer graphene powder (Fig. S9). The saturation in $\rho_{HD}$ at 600 °C occurred after a few hours and was below 50%, indicating poisoning of the reference catalysts (Fig. 3*B*). Next, to assess a possible role played in the production of HD by various atomic-scale defects in graphene such as vacancies and edges as well as by residual oxygen and other functional groups present in the graphene powder (28), we have compared it with other possible carbon-based catalysts (namely, few-layer graphene and activated charcoal). Both showed substantially lower catalytic activity for the exchange reaction (Fig. 3*B*). Our few-layer graphene used in the control experiments was a powder obtained by ultrasonic exfoliation (31). Its flakes were similar in size to those in the monolayer graphene powder (several micrometers) but considerably thicker (typically, 10 layers). They were notably less corrugated (Fig. S8), as expected (20-23). Few-layer graphene showed much less efficiency than monolayer graphene not only in terms of HD production per g but also per surface area (~30 times less efficient per m$^2$; see Fig. S9*D*). This agrees well with the fact that few-layer graphene has a much less rippled surface than monolayer graphene. Furthermore, activated charcoal known for its highly porous structure provided numerous broken carbon bonds and oxygen surface groups (32). Nonetheless, its catalytic efficiency for HD production was only minor, confirming that defects (possibly present in minor amounts) play little role in the observed hydrogen dissociation by graphene. The above comparison between different graphitic materials corroborates that nanoripples – which are unavoidable for graphene monolayers, less profound for few-layer graphene and absent in charcoal – play an important role in the observed catalytic activity.

To compare catalytic efficiency of the tested materials more quantitatively, we also translated their $\rho_{HD}$ at saturation (Fig. 3*B*) into the number of HD molecules, $N_{HD}$, produced per gram or surface area of the catalysts. Fig. S9*C* shows that, in terms of the weight catalytic efficiency ($N_{HD}$ per g), graphene excels all the other powders by orders of magnitude. This is helped by the fact that graphene is essentially a surface and has no bulk (theoretical surface area of ~2,600 m$^2$ g$^{-1}$) whereas the other powders have their bulk mostly inaccessible for the HD production. Even in terms of the surface efficiency ($N_{HD}$ per m$^2$), graphene compares favorably with MgO and ZrO$_2$ and is surpassed only by Cu (Fig. S9*D*). From the known $N_{HD}$, we can estimate the turnover number *TON* defined as the maximum yield attainable from a catalytic center (33). Assuming that all carbon atoms of graphene contribute equally to the reactivity, we find *TON* $\approx$ 0.1. On the other hand, the D peak intensity in the Raman spectra of graphene in Fig. 2*A* indicates (24) that only 8-10% of graphene's atoms took part in the reaction before the material got poisoned. Those were presumably carbon atoms located in the most curved part of graphene where the activation energy was strongly reduced (Fig. 1*F*). Accounting only for the latter atoms as catalytic centers, we obtain *TON* $\approx$ 1. Finally, if we consider individual nanoripples as catalytic centers, this yields huge *TON* $\approx$ 10$^3$–10$^4$. Note that, from the perspective of traditional catalysis, these estimates have only a limited sense because of the difficulty with defining what catalytic centers are in our case. In particular, carbon atoms on dynamic ripples may change in time from being reactive to non-reactive and back. However, even our most conservative estimate of *TON* $\approx$ 1 makes graphene competitive with traditional catalysts because in the latter case only near-surface atoms play a role in stimulating reactions whereas most of graphene's atoms partake in nanorippling and thus are reactive.

**Discussion**

A flat sheet of graphene is expected to be highly stable and chemically inert under ambient conditions, similar to its parent material, graphite. However, graphene monolayers are never perfectly flat because of thermal fluctuations (flexural phonons) and practically unavoidable local strain, which generate static nanoscale ripples (20-23). Our experiments show that in terms of reactivity such nanorippled graphene is quite different from both graphite and atomically flat graphene. This offers strong support for earlier theoretical studies of hydrogen dissociation on curved graphene (14, 17-19) and, also, indicates that nanoripples can be more important for catalysis than the 'usual suspects' such as, for example, vacancies, edges and residual functional groups on graphene's surfaces. To enhance graphene's catalytic activity, one can increase the density of nanoripples by increasing *T* that induces thermal fluctuations and creates dynamic ripples or by placing graphene onto rough substrates (static ripples). Furthermore, additional ripples can be formed by thermal cycling of graphene on various substrates due to different thermal expansion coefficients. Because nanorippling is inherent for all atomically thin crystals, it is worth keeping in mind that the enhancement of chemical and/or catalytic activity is also possible for the case of other 2D materials and for other chemical reactions. For example, bulk $MoS_2$ and other chalcogenides are often used as 3D catalysts but may exhibit even stronger activity in the 2D form.

Our results have implications for many previous observations reported in the literature. For example, nanoripples can be relevant for graphene oxidation that occurs preferentially on monolayers and wrinkles but has remained unexplained (8, 11). To this end, note that monolayer hBN (no activity with respect to hydrogen splitting, as found in our work) also exhibits much stronger thermal oxidation resistance than graphene (34, 35). This is consistent with their different electronic spectra that are ultimately responsible for the reported difference in hydrogen dissociation on nanoripples. Our results also support the idea of hydrogen storage inside carbon nanotubes (17). Indeed, molecular hydrogen can adsorb and dissociate on strongly curved graphene surfaces, then flip through as protons and finally desorb from the inner concave surface so that $H_2$ can potentially be stored inside carbon nanotubes. Furthermore, in heterogeneous catalytic reactions involving, for example, graphene-coated metal surfaces and nanoparticles (36-40), the local curvature of graphene can potentially account for some enhanced reactivity, a mechanism disregarded so far. The hydrogen dissociation on nanoripples may also play a role in other reactions involving graphene-based catalysts (e.g., in electrolysis and electrocatalysis) (12, 36-40).

1. B. F. Machado, P. Serp, *Catal. Sci. Technol.* **2**, 54–75 (2012).
2. S. Navalon, A. Dhakshinamoorthy, M. Alvaro, H. Garcia, *Chem. Rev.* **114**, 6179–6212 (2014).
3. X. B. Fan, G. L. Zhang, F. B. Zhang, *Chem. Soc. Rev.* **44**, 3023–3035 (2015).
4. D. Deng *et al.*, *Nat. Nanotechnol.* **11**, 218–230 (2016).
5. E.J. Askins *et al.*, *Nat. Commun.* **12**, 3288 (2021).
6. D. R. Dreyer, H.-P. Jia, C. W. Bielawski, *Angew. Chem. Int. Ed.* **49**, 6813–6816 (2010).
7. A. Primo *et al.*, *Nat. Commun.* **5**, 5291 (2014).
8. L. Liu *et al.*, *Nano Lett.* 8, 1965–1970 (2008).
9. Q. Wu *et al.*, *Chem. Commun.* **49**, 677–679 (2013).
10. M. A. Bissett, S. Konabe, S. Okada, M. Tsuji, H. Ago, *ACS Nano* **7**, 10335–10343 (2013).
11. Y. H. Zhang *et al.*, *Carbon* **70**, 81–86 (2014).
12. T. Kosmala *et al.*, *Nat. Catal.* **4**, 850–859 (2021).
13. F. Li *et al., Nat. Commun.* **13**, 4472 (2022).
14. P. Z. Sun *et al.*, *Nature* **579**, 229–232 (2020).
15. S. Hu *et al.*, *Nature* **516**, 227–230 (2014).
16. M. Lozada-Hidalgo *et al.*, *Science* **351**, 68–70 (2016).
17. S. M. Lee, K. H. An, Y. H. Lee, G. Seifert, T. Frauenheim, *J. Am. Chem. Soc.* **123**, 5059–5063 (2001).
18. D. W. Boukhvalov, M. I. Katsnelson, *J. Phys. Chem. C* **113**, 14176–14178 (2009).
19. H. McKay, D. J. Wales, S. J. Jenkins, J. A. Verges, P. L. de Andres, *Phys. Rev. B* **81**, 075425 (2010).
20. J. C. Meyer *et al.*, *Solid State Commun.* **143**, 101–109 (2007).
21. A. Fasolino, J. H. Los, M. I. Katsnelson, *Nat. Mater.* **6**, 858–861 (2007).
22. R. Zan *et al.*, *Nanoscale* **4**, 3065–3068 (2012).
23. P. Xu *et al.*, *Nat. Commun.* **5**, 3720 (2014).
24. D. C. Elias *et al.*, *Science* **323**, 610–613 (2009).
25. K. E. Whitener, *J. Vac. Sci. Technol.* A **36**, 05G401 (2018).

26. P. Z. Sun *et al.*, *Nat. Commun.* **12**, 7170 (2021).
27. Z. Ni, Y. Wang, T. Yu, Z. Shen, *Nano Res.* **1**, 273–291 (2008).
28. J. A. Hondred, L. R. Stromberg, C. L. Mosher, J. C. Claussen, *ACS Nano* **11**, 9836–9845 (2017).
29. G. C. Bond, (Academic, New York, 1962).
30. D. A. Dowden, N. Mackenzie, B. M. V. Trapnell, *Adv. Catal.* **9**, 65–69 (1957).
31. Y. Hernandez *et al.*, *Nat. Nanotechnol.* **3**, 563–568 (2008).
32. F. Rodríguez-Reinoso, *Carbon* **36**, 159–175 (1998).
33. S. Kozuch, J. M. L. Martin, *ACS Catal.* **2**, 2787–2794 (2012).
34. L. H. Li, J. Cervenka, K. Watanabe, T. Taniguchi, Y. Chen, *ACS Nano* **8**, 1457–1462 (2014).
35. L. Shen *et al.*, *J. Mater. Chem. A* **4**, 5044–5050 (2016).
36. M. Tavakkoli *et al.*, *Angew. Chem. Int. Ed.* **54**, 4535–4538 (2015).
37. J. Deng, P. Ren, D. Deng, X. Bao, *Angew. Chem. Int. Ed.* **54**, 2100–2104 (2015).
38. T. Sharifi *et al.*, *Carbon* **141**, 266–273 (2019).
39. G. Cilpa-karhu, O. J. Pakkanen, K. Laasonen, *J. Phys. Chem. C* **123**, 13569–13577 (2019).
40. K. Hu *et al.*, *Nat. Commun.* **12**, 203 (2021).

**Supporting Information**

**1. Microcontainers for hydrogen permeation experiments.** To fabricate such containers, we followed the procedures developed in ref. (1). Monocrystals of graphite with a thickness of 150–200 nm were mechanically exfoliated onto an oxidized silicon wafer. The crystals' surface was carefully examined in an optical microscope using both dark-field and differential-interference-contrast modes. Areas free from contamination and atomic terraces were selected for the next step that involved electron-beam lithography. It was employed to make a polymer mask defining an array of rings with inner diameters of 0.5–1 µm and rims having a width of ∼1 µm. Dry etching was used to project the mask geometry into the graphite crystals, which created microwells having ∼80 nm depth (1). After dissolving the polymer mask, the structures were annealed at 400 °C in a $H_2$/Ar atmosphere for several hours to remove polymer residues. Next, large (> 100 µm in size) graphene or hBN monolayers were obtained by mechanical exfoliation and transferred on top of the microwells to seal them, creating microcontainers (Fig. 1*B*).
The resulting microcontainers were first inspected by AFM (*Fastscan* from *Bruker*) for any damage in the atomically tight sealing (1) and possible defects in suspended membranes (e.g., tears, cracks and wrinkles). Only microcontainers without discernible imperfections progressed to the next stage that was leakage tests. First, microcontainers were placed inside a stainless-steel chamber that was pressurized with a heavy inert gas (for example, Ar or Xe under a typical pressure of 3 bar). After being stored for one week, the microcontainers were quickly taken out and, within a few minutes, measured by AFM to find possible changes in the membrane position $\delta$ (Figs. 1*B*, *C*). Such changes would indicate the presence of atomic-scale defects allowing gas permeation inside microcontainers (1-4). Only those exhibiting no changes in $\delta$ (within our experimental accuracy of better than 1 nm; Fig. 1*D*) were further tested by placing them into a 1-bar helium atmosphere for one month. This test ensured that even smallest defects such as individual vacancies were absent (2), that is, the microcontainers were perfectly sealed and defect free (1). These helium-tight devices were used in the hydrogen permeation measurements described in the main text.

**2. Leak tests using silicon-oxide microcontainers.** The above tests could easily detect atomic-scale defects (2) but thousand times larger defects that allowed very rapid deflation of microcontainers (quicker than within a few minutes) could not be revealed using the described approach. Indeed, no changes in $\delta$ would be detected not only for He-tight microcontainers but also those with large cracks in membranes. To rule out the latter possibility, we mostly relied on dedicated atomic force and scanning electron microscopy (1, 2). Nonetheless, as an additional proof that large defects were generally absent in our graphene and hBN membranes, we tested microcontainers made from oxidized silicon wafers, as first explored in refs. (3, 4). The difference is that amorphous $SiO_2$ does not provide good sealing but allows a slow gas permeation through it (3, 4), unlike our monocrystalline containers with atomically-tight sealing (1). Therefore, if one compares similarly made and sealed silicon-oxide and monocrystalline microcontainers, only the former should exhibit notable bulging in Ar and He (3, 4).
With this test in mind, we etched microwells in an oxidized silicon wafer (300 nm of $SiO_2$) and sealed them with exfoliated graphene or hBN monolayers (Fig. S1*A*). The microcontainers were typically ∼3 µm in diameter and ∼200 nm deep. Permeation of various inert gases into such microcontainers was measured following the procedures described above and in refs. (1-4). Taking monolayer hBN and Ar as an example in Fig. S1*B*, the suspended membranes clearly bulged out after their pressurization. Their consecutive deflation in air was monitored by AFM as a function of time, yielding a deflation rate of ~0.7 nm/h (Fig. S1*B*). This translates into a permeance of ~2 × $10^{-27}$ mol $s^{-1}$ $Pa^{-1}$ for Ar, which is consistent with the previously reported values for $SiO_2$ microcontainers (3, 4). Similar agreement was also observed for other gases including He. This corroborates the previous conclusions that the leakage occurs through $SiO_2$ rather than monocrystalline membranes. Our extensive leakage tests unambiguously prove that the membranes made from exfoliated graphene and hBN monolayers were completely free from either microscopic or macroscopic defects. Accidental large defects were ruled out by AFM and electron microscopy.

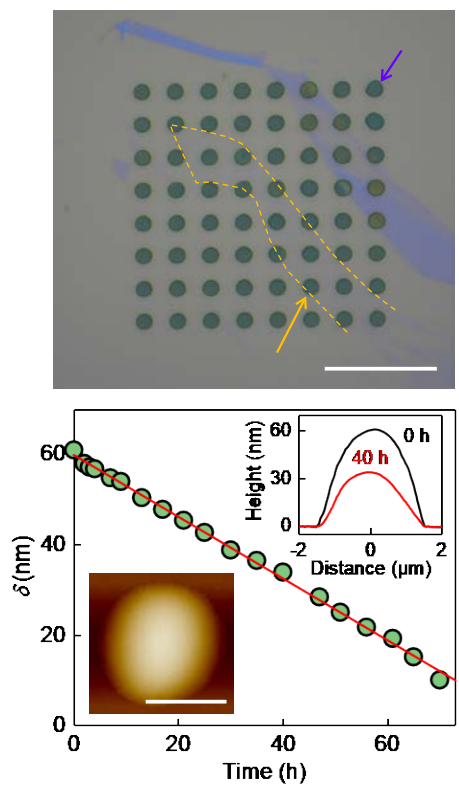

**Fig. S1.** Silicon-oxide microcontainers. (*A*) Optical micrograph of an array of microwells etched in silicon oxide and sealed with monolayer hBN. The dashed curve outlines the monolayer position. The blueish regions are thicker hBN. (*B*) Deflation for a representative $SiO_2$ microcontainer after its storage under 2 bar of Ar for 10 days. Lower inset: AFM image after the pressurization. Color scale (dark-to-bright), 0 to 60 nm. Scale bar, 2 μm. Upper inset: height profiles acquired within minutes after the microcontainer was taken out of a pressure chamber and after 40 hours (color coded). All the AFM measurements were done in air at room temperature.

**3. Density functional theory (DFT) calculations.** Dissociation of molecular hydrogen on graphene and hBN nanoripples was simulated using DFT, as implemented in Vienna *ab initio* package (5). The ion-electron interactions and exchange correlation potential were described using the projected augmented wave and generalized gradient approximation (6). The kinetic energy cutoff and *k*-point meshes were set at 500 eV and 3×3×1, respectively (7). To avoid periodic interactions, a vacuum region of 20 Å was adopted. The convergences for the stress force and total energy were set as 0.02 eV/Å and $10^{-5}$ eV, respectively. The van der Waals interactions of $H_2$ with graphene and hBN were treated by the semi-empirical DFT-D3 method (8, 9). The ripples were characterized by the ratio $t/D$ of their height $t$ to the corrugation diameter $D$ (inset of Fig. 1*F*). Corrugated supercells comprising 8×8 graphene or hBN unit cells with a nonzero $t/D$ were created by applying biaxial compression. The initial states were constructed using $H_2$ being physiosorbed on the surface. Then, the $H_2$ molecule was allowed to undergo dissociation until a final state was reached in which two hydrogen adatoms were chemisorbed at specific locations. The electron distribution was optimized during the reaction process. The energy barrier $E_b$ for the reaction pathway was calculated using the climbing-image nudged elastic band (NEB) method (10). All atoms were allowed to fully relax to the ground states and the spin polarization was also taken into account.

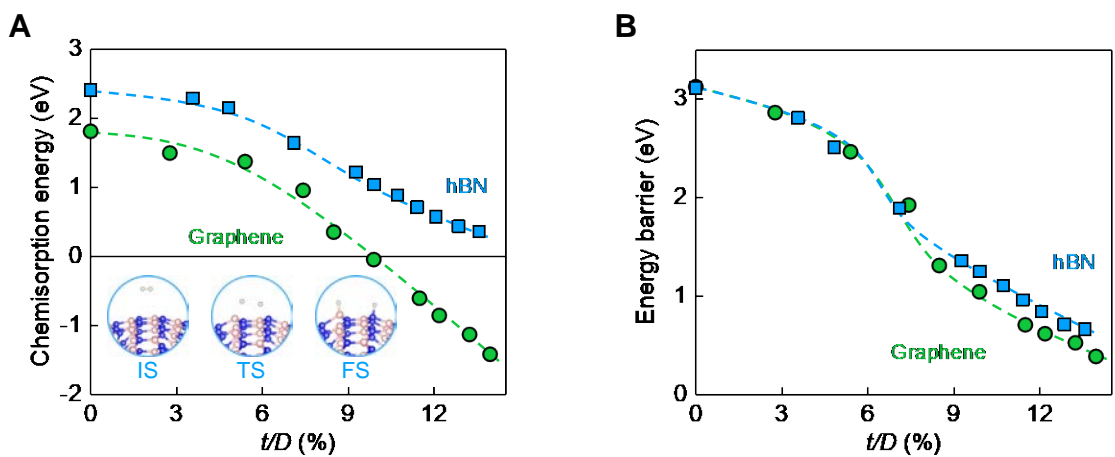

**Fig. S2.** Hydrogen dissociation at graphene and hBN ripples. (*A*) Chemisorption energy $E_c$ and (*B*) energy barrier $E_b$ as a function of $t/D$ for graphene and hBN (color coded). The hydrogen adatoms are adsorbed at the central positions illustrated in the insets of Fig. 1*F*. Insets in (*A*): schematics of the initial (IS), transitional (TS) and final states (FS) for this process in the case of hBN ripples. Dashed curves: guides to the eye.

Fig. S2 shows evolution of $E_b$ and the chemisorption energy $E_c$ for hydrogen dissociation on graphene ripples with increasing their $t/D$. The final state constitutes two H adatoms being adsorbed at the central positions of the graphene hexagon (inset of Fig. 1*F*). We found this configuration to be most energetically favorable for hydrogen dissociation, after trying many different adsorption positions and reaction pathways. In agreement with the earlier report (1), the dissociation reaction becomes energetically favorable (that is, exothermic with $E_c < 0$) only for ripples with $t/D$ larger than ~10%, and the barrier $E_b$ also decreases to < 1 eV for such ripples. A further increase in the curvature leads to a rapid decrease in the energy barrier. For example, $E_b \approx 0.4$ eV is attained at $t/D \approx 13\%$ (Fig. S2*B*). Such nanoripples were visualized on suspended graphene membranes by high-resolution electron and tunnelling microscopies (11-13). The inferred $E_b$ is also consistent with the reaction activation energy estimated from our experiments (Figs. 2 and 3). The above analysis supports the concept of strongly-curved nanoripples as the reactive sites for hydrogen dissociation on graphene.

Our DFT analysis for hBN monolayers revealed a radically different behavior. The central position for similar hBN ripples is energetically unfavorable ($E_c > 0$) for the whole range of $t/D$ considered in our calculations (Fig. S2A). Larger curvatures (> 15%) are unrealistic for hBN because they would eventually break the crystal lattice. In addition to the central position, we have carried out DFT simulations for H atoms being adsorbed at other positions including the bridge B-N, nearest N-N and nearest B-B positions (insets of Fig. S3C). In all cases, we found the hydrogen splitting to be energetically unfavorable. This shows that, unlike nanorippled graphene, nanorippled hBN is highly inert, at least with respect to hydrogen dissociation. In addition, we analyzed the effect of pure strain on dissociation of $H_2$ by considering biaxial tensile strain for a flat hBN lattice (Fig. S3). Same as in the case of hBN ripples, the reaction was found energetically unfavorable ($E_c > 0$) for all positions and reasonable strains considered (Fig. S3).

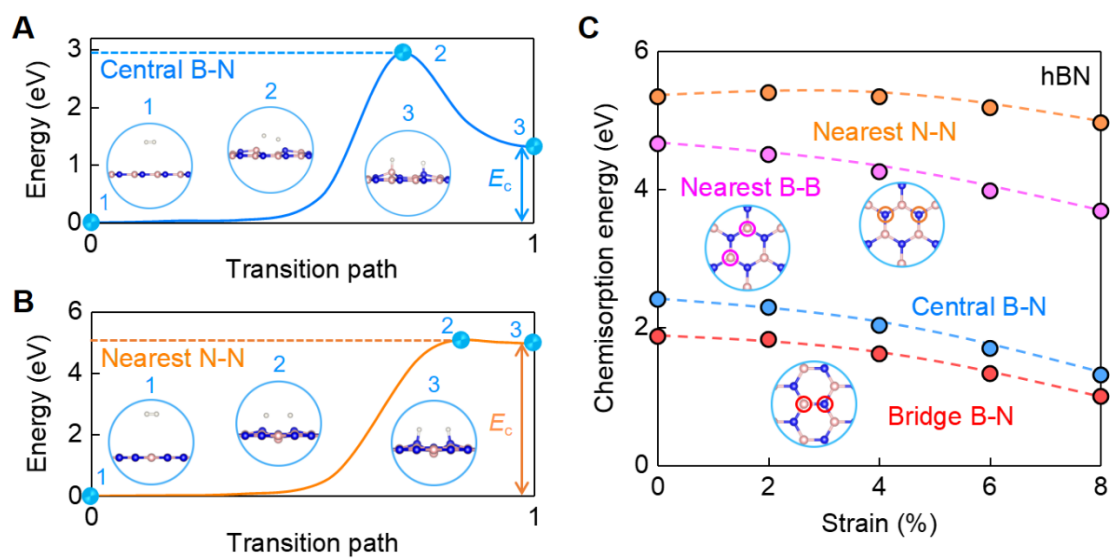

**Fig. S3.** Hydrogen dissociation on hBN under strain. Energy profiles for hydrogen dissociation on hBN with two H adatoms being adsorbed at (*A*) central B-N and (*B*) nearest N-N positions. A biaxial tensile strain of 8% is applied to the flat hBN lattice. Insets in (*A*, *B*) schematically show the atomic configurations of IS, TS and FS, respectively. (*C*) Chemisorption energy $E_c$ versus strain for different adsorption positions (color coded and illustrated in the insets). Dashed curves: guides to the eye.

**4. Different reactivity of graphene and hBN, according to DFT analysis.** The discussed splitting of $H_2$ on graphene and hBN surfaces involves the breaking of H-H bonds in the dissociated molecule, resulting in H atoms, and their bonding to the crystal lattice to form C-H or B/N-H bonds. Accordingly, the energy evolution of this reaction should be determined by the lengths and strengths of the involved bonds. Trying to understand the different reactivities of graphene and hBN ripples, we performed two additional calculations.

First, we analyzed the evolution of bond lengths (namely, H-H bonds and C-H or B/N-H bonds) along the hydrogen dissociation pathway. Fig. S4 shows the potential energy surface (PES) for graphene and hBN ripples as a function of the bond length between two dissociated hydrogen atoms ($d_{H-H}$) and their distance to the lattice ($d_{H-C}$ or $d_{H-B/N}$). According to the shape of the saddle points, the transition states (TS) identify the evolution of the pathway profiles from the initial states (IS) to the final states (FS). By comparing the bond lengths at TS, we have found that dissociation of H₂ on graphene ripples requires a shorter elongation of H-H bonds than that on hBN ripples and, simultaneously, the former reaction can happen at a longer distance from the surface. Therefore, the energy cost for the reaction of H₂ with graphene ripples should be notably smaller than that with hBN ripples, as clearly revealed by the PES.

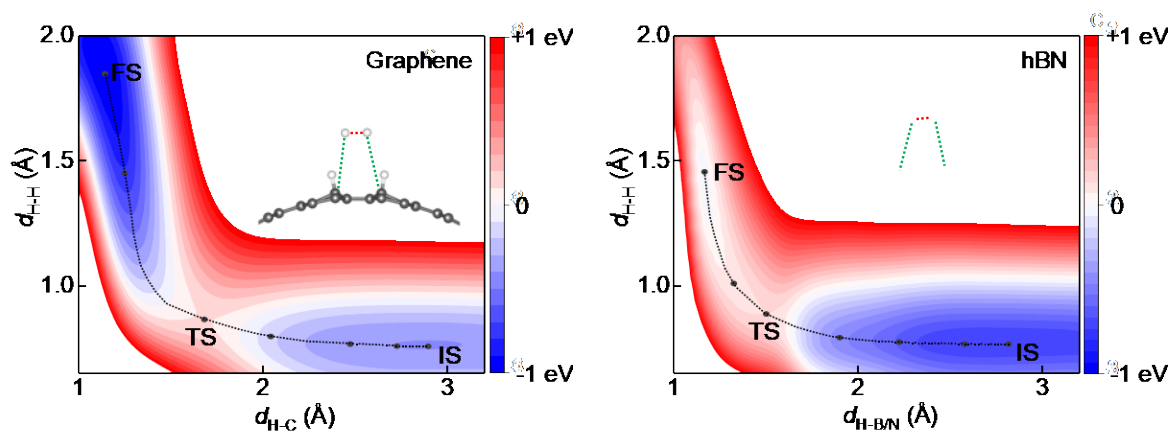

**Fig. S4.** Potential energy surface (PES) and changes in bond lengths. The PES for hydrogen dissociation at the central position for (*A*) graphene and (*B*) hBN ripples as a function of the distance between two dissociated H atoms ($d_{H-H}$), and their distance to the lattice ($d_{H-C}$ or $d_{H-B/N}$). In both cases, $t/D$ = 12%. The black dashed curves indicate the minimum energy paths. Insets: schematics of the dissociation process.

Second, we analyzed bonding strengths of atomic hydrogen to graphene and hBN ripples. To this end, we used the crystal orbital Hamilton population (COHP) method (14). As shown in Fig. S5, the negative and positive COHP values correspond to the antibonding and bonding states, respectively. For graphene ripples (Fig. S5*A*), the occupied states are all bonding states with energy distributed between -10 eV and -5 eV, below the Fermi level ($E_f$) whereas the unoccupied states are all antibonding and located at least 3 eV above $E_f$. This configuration indicates a stable adsorption of atomic H on graphene ripples. In contrast, hBN ripples exhibit a notable antibonding component in the occupied states at energies between 0 and -3 eV (Fig. S5*B*). Accordingly, the total energy is increased, resulting in a less stable adsorption configuration. This analysis provides an additional understanding of the difference between reactivities of graphene and hBN ripples.

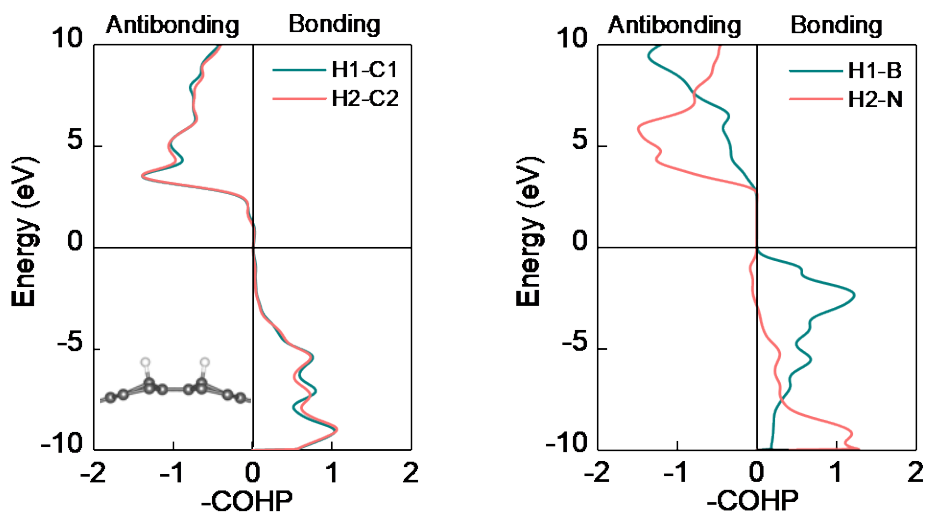

**Fig. S5.** Crystal orbital Hamilton population (COHP) bonding analysis. Interaction of hydrogen adatoms with (*A*) graphene and (*B*) hBN ripples ($t/D$ = 12%). Insets: illustration of atomic structures for hydrogen adatoms adsorbed at the central position for graphene and hBN ripples.

**5. Raman measurements.** Our monolayer graphene crystals were either exfoliated onto an oxidized silicon wafer (RMS ≈ 0.5 nm) or transferred onto an atomically flat surface of graphite (Fig. S6). In the latter case, crystallographic axes of graphene and graphite were intentionally non-aligned. We tried other atomically flat surfaces obtained by exfoliation of layered materials but all of them exhibited a large background in the spectral region of graphene's D peak.

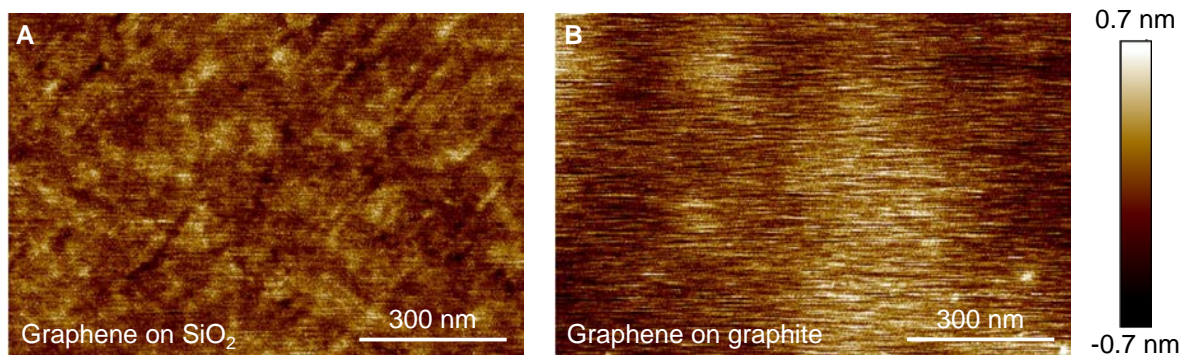

**Fig. S6.** AFM images of graphene on different substrates. Exfoliated monolayer crystals were deposited onto (*A*) a silicon oxide wafer and (*B*) an atomically flat surface of graphite. Same color scale (-0.7 nm to +0.7 nm) for both images. Nanoscale local corrugations are evident for graphene on SiO$_2$ but indiscernible above noise for graphene on graphite (only smooth variations in height are visible in the latter case).

For Raman analysis, graphene samples were placed into a vacuum-tight heating stage with a quartz optical window (*Linkam*). The stage was employed in combination with Raman microscope *WITec*. During the measurements, a continuous flow of either hydrogen or helium was provided into the stage through dedicated gas in- and out- lets. If repeated measurements were necessary to acquire temperature dependences or check the effect of different gases, Raman spectra were acquired for the same spots far away from graphene edges.

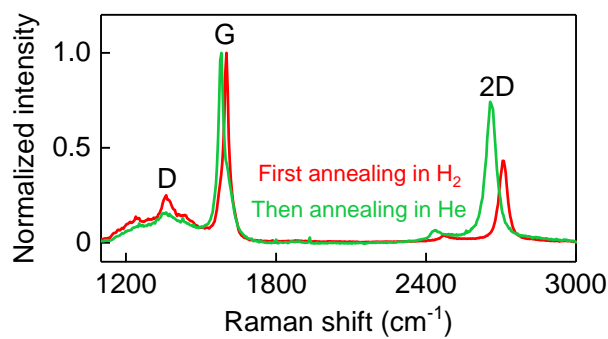

**Fig. S7.** Changes in Raman spectra of hydrogenated graphene after its annealing in helium. The spectra were taken at room $T$ from the same position and normalized to have the same amplitude of the G peak. Red curve: after the initial exposure to 1-bar hydrogen at 600 °C for 2 hours. The developed D peak could be partially annealed using a helium atmosphere (green curve; 600 °C for 8 h). Same laser parameters as specified in Fig. 2 of the main text.

To illustrate how annealing in a helium atmosphere influenced Raman spectra of hydrogenated graphene, we provide Fig. S7. This is in addition to the data points in the inset of Fig. 2*B*. The D peak developed due to heat-treatment of graphene in hydrogen was clearly reduced after its exposure to the same temperature but in helium. Annealing for longer times in either helium or vacuum led to a further decrease in the D peak amplitude but a residual hump in the D peak region remained even after several days. This can be attributed to the formation of stable regions of graphane, in agreement with the previous report (15).

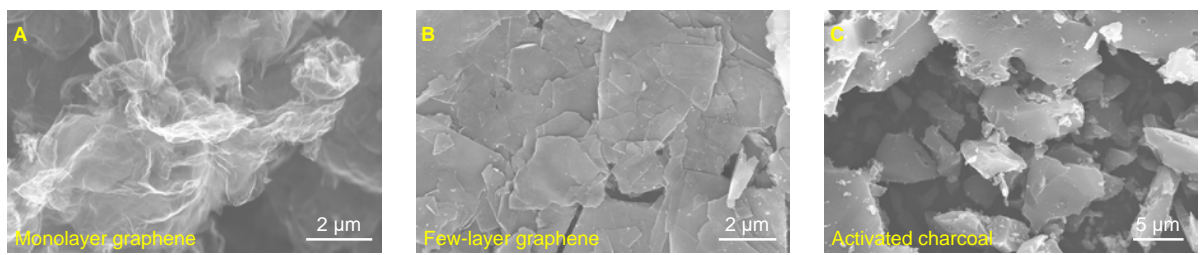

**Fig. S8.** Electron micrographs of the three carbon-based materials tested as catalysts. (*A*) Monolayer graphene powder. (*B*) Few-layer graphene powder. (*C*) Activated charcoal.

**6. Details of hydrogen isotope exchange reaction.** Monolayer graphene powder (*ACS Material*, specific surface area ~1000 m$^2$ g$^{-1}$) was used as a catalyst for dissociation-recombination reaction $H_2 + D_2 \leftrightarrow 2HD$ as described in the main text. Detailed characterization of the powder is provided in ref. (16). For comparison, we tested other carbon-based materials as possible catalysts, namely, few-layer graphene obtained by ultrasonic exfoliation (17) (typical thickness of 10 layers; specific surface area of 100–200 m$^2$ g$^{-1}$) and activated charcoal (*J. L. BRAGG'S*, specific surface area of ~1000 m$^2$ g$^{-1}$). Fig. S8 shows morphologies of the three carbon-based materials. As a reference, we also employed the known catalysts for the HD reaction (18, 19): $ZrO_2$ (particle size, ~5 μm; specific surface area, ~0.2 $m^2$ $g^{-1}$), MgO (particle size, ≤ 50 nm; specific surface area, ~30 $m^2$ $g^{-1}$) and Cu (particle size, ~50 μm, specific surface area, ~0.01 $m^2$ $g^{-1}$). The latter materials were acquired from *Sigma-Aldrich*.

To assess their catalytic efficiency, the entire volume of a quartz tube (length of 300 mm; inner diameter of 5 mm) was filled up with one of the tested materials. The powders were loosely packed allowing easy access of gases throughout the tube and its vacuumization. Because of large differences in densities and packing, the weight of the powders placed inside the tube varied considerably. To be specific, it required ~10 mg of the monolayer graphene powder, ~200 mg of the few-layer one, ~1 g of activated charcoal and 2 to 10 g of the $ZrO_2$, MgO and Cu powders. The tube was then sealed and pumped down to ~$10^{-3}$ mbar. A 50% $H_2$ – 50% $D_2$ mixture with the total pressure $P ≈ 1$ bar was put into the chamber and heat-treated at a chosen temperature (up to 600 °C) for a specific length of time.

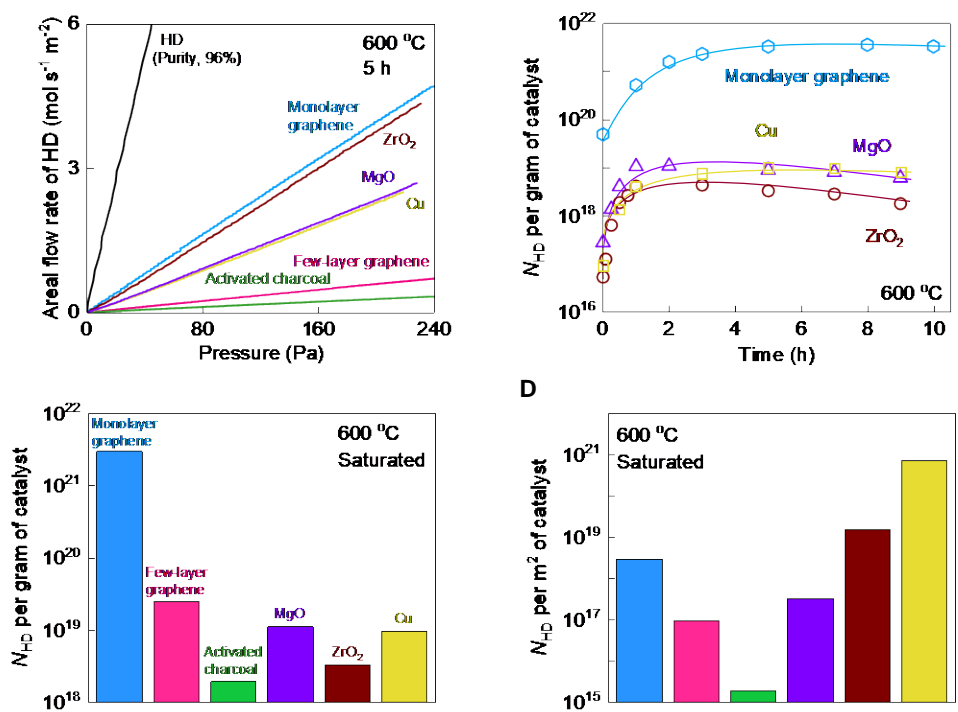

**Fig. S9.** Determining HD in the heat-treated hydrogen-deuterium mixtures. (*A*) Flow rates of HD through a micrometer aperture as a function of the feed pressure. In this case, a 50% $H_2$ – 50% $D_2$ mixture was annealed at 600 °C for 5 h using different catalysts (color coded). (*B*) HD production as a function of time for $ZrO_2$, MgO, Cu and graphene (color coded). Solid curves: guides to the eye. (*C*) Total amount of the produced HD ($N_{HD}$ per gram) after the saturation was reached for all the tested materials (5 h at 600 °C). (*D*) Same as in panel (*C*) but recalculated in terms of the surface area.

For mass spectrometry measurements, the heat-treated gas was allowed to flow through an aperture of the known diameter (typically, 6 μm). The aperture was made within a freestanding silicon-nitride membrane (500 nm thick). HD flow rates at different feed pressures were measured by a leak detector that allowed measurements for masses 2, 3 and 4 (*Leybold*). The spectrometer was calibrated using commercially supplied HD, $D_2$ and $H_2$ gases (*Sigma-Aldrich*). Fig. S9*A* shows examples of our measurements after heat-treating the $H_2 - D_2$ mixture with various catalysts under the same conditions (600 °C for 5 h). From these curves, the saturation HD concentrations $\rho_{HD}$ could be calculated as done in Fig. 3*B* of the main text.

We have also measured the HD production as a function of time for the reference catalysts. They exhibited kinetic behaviors similar to that of the monolayer graphene powder (Fig. S9*B*). The saturation in HD production was typically observed after 1 h at 600 °C for $ZrO_2$ and MgO, whereas it required ~5 h for Cu, similar to the case of graphene. Figs. S9*C-D* compare the total numbers of HD molecules, $N_{HD}$, produced using the different standard catalysts and carbon-based materials. The $N_{HD}$ values were normalized with respect to both weight and surface area.

1. P. Z. Sun *et al.*, *Nature* **579**, 229–232 (2020).
2. P. Z. Sun *et al.*, *Nat. Commun.* **12**, 7170 (2021).
3. J. S. Bunch *et al.*, *Nano Lett.* **8**, 2458–2462 (2008).
4. S. P. Koenig, L. Wang, J. Pellegrino, J. S. Bunch, *Nat. Nanotechnol.* **7**, 728–732 (2012).
5. G. Kresse, J. Furthmuller, *Phys. Rev. B* **54**, 11169–11186 (1996).
6. J. P Perdew, K. Burke, M. M. Ernzerhof, *Phys. Rev. Lett.* **77**, 3865–3868 (1996).
7. H. J. Monkhorst, J. D. Pack, *Phys. Rev. B* **13**, 5188–5192 (1976).
8. S. Grimme, *J. Comput. Chem.* **27**, 1787–1799 (2006).
9. T. Kerber, M. Sierka, J. Sauer, *J. Comput. Chem.* **29**, 2088–2097 (2008).
10. D. Sheppard, P. Xiao, W. Chemelewski, D. D. Johnson, G. Henkelman, *J. Chem. Phys.* **136**, 074103 (2012).
11. J. C. Meyer *et al.*, *Solid State Commun.* **143**, 101–109 (2007).
12. R. Zan *et al.*, *Nanoscale* **4**, 3065–3068 (2012).
13. P. Xu *et al.*, *Nat. Commun.* **5**, 3720 (2014).
14. V. L. Deringer, A. L. Tchougréeff, R. Dronskowski, *J. Phys. Chem. A* **115**, 5461–5466 (2011).
15. D. C. Elias *et al.*, *Science* **323**, 610–613 (2009).
16. J. A. Hondred, L. R. Stromberg, C. L. Mosher, J. C. Claussen, *ACS Nano* **11**, 9836–9845 (2017).
17. Y. Hernandez *et al.*, *Nat. Nanotechnol.* **3**, 563–568 (2008).
18. G. C. Bond, (Academic, New York, 1962).
19. D. A. Dowden, N. Mackenzie, B. M. V. Trapnell, *Adv. Catal.* **9**, 65–69 (1957).